\documentclass[a4paper,prb,aps,superscriptaddress,preprint,unsortedaddress,fleqn]{revtex4}
\usepackage{graphicx}
\usepackage{amsmath}
\usepackage{amssymb}
\usepackage{amsfonts}
\usepackage{indentfirst}
\usepackage[T1]{fontenc}
\usepackage[latin1]{inputenc}

\begin{document}

\title{Rigorous approach to the nonlinear saturation of the tearing mode in cylindrical and slab geometry}
\author{N. Arcis}
\affiliation{Association Euratom-CEA, DSM/DRFC/SCCP, CEA
Cadarache, b\^{a}t. 513, 13108 St-Paul-lez-Durance, France}
\author{D.F. Escande}
\affiliation{UMR 6633 CNRS-Universit\'{e} de Provence, case 321,
Centre de St-J\'{e}r\^{o}me, 13397 Marseille cedex 20, France}
\author{M. Ottaviani}
\affiliation{Association Euratom-CEA, DSM/DRFC/SCCP, CEA
Cadarache, b\^{a}t. 513, 13108 St-Paul-lez-Durance, France}

\begin{abstract}
The saturation of the tearing mode instability is described within the standard framework of reduced magnetohydrodynamics (RMHD) in the 
case of an $r$-dependent or of a uniform resistivity profile. Using the technique of matched asymptotic expansions, where the perturbation 
parameter is the island width $w$, the problem can be solved in two ways: with the so-called flux coordinate method, which is based on the 
fact that the current profile is a flux function, and with a new perturbative method that does not use this property. The latter is applicable to 
more general situations where an external forcing or a sheared velocity profile are involved. The calculation provides a new relationship 
between the saturated island width and the $\Delta '$ stability parameter that involves a $\ln{w/w_{0}}$ term, where $w_{0}$ is a 
nonlinear scaling length that was missing in previous work. It also yields the modification of the equilibrium magnetic flux function.

\end{abstract}

\maketitle

\newpage

\section{INTRODUCTION}

In magnetized plasmas with current inhomogeneities, magnetic islands can develop as the consequence of tearing mode instabilities. 
This mode corresponds to a global magnetic perturbation that is resonant on a magnetic surface where its wave-number is perpendicular to 
the magnetic field. From a theoretical viewpoint, the instability is allowed by non-ideal plasma effects such as resistivity \cite{Furth}, 
electron inertia \cite{Coppi} and more generally kinetic effects \cite{LP,LPV}. On the resonance surface, there is a modulated current 
sheet whose  current density is proportional to the classical tearing mode stability parameter $\Delta'$. This parameter, whose precise 
definition is reviewed below, depends solely on the ideal magnetohydrodynamics (MHD) properties of the system. Instability occurs when 
$\Delta'$ exceeds a critical value, ($\Delta' > \Delta_c \geq 0$). This value depends on the actual physics included in the model. For 
instance, in visco-resistive MHD, $\Delta_c$ is a function of viscosity and resistivity \cite{Tebaldi,Dahlburg}. Above the critical value, the 
resonant magnetic surface can break up (tear) and is substituted by a chain of topologically distinct structures called  \emph{magnetic islands}. 
Until recently, tokamak operation avoided the formation of such islands, since they result in greater radial transport and hence deteriorate 
particle and energy confinement. However, in certain experimental conditions, a magnetic island may help the plasma  form a stable internal 
transport barrier \cite{Luce}. In the reversed-field pinch (RFP), the occurrence of several magnetic island chains leads to magnetic chaos 
with reduced confinement, but the formation of a single chain is desirable, since it should provide good magnetic flux surfaces and a laminar 
dynamo. Therefore, a correct description and understanding of the nonlinear tearing mode is both important for thermonuclear fusion and for 
advancing the theory of plasma self-organization.

In the theory of magnetic island formation, nonlinear effects come into play as soon as the island width exceeds the width of the boundary 
layer at the resonant surface, as given by linear theory. The nonlinear tearing mode is classically described by applying  resistive reduced  
magnetohydrodynamics \cite{Strauss} (RRMHD) to the model of a static plasma slab, in the limit of small dissipation. The magnetic island 
region is considered as a boundary layer whose nonlinear features are dealt with,  while the outer region is adequately described by linear 
theory only. The inner and outer solutions are then matched asymptotically. Rutherford \cite{Ruth} showed that the island growth is 
sufficiently slow that inertia can be neglected in the inner solution. The fluid equations of RRMHD then reduce to a mere force balance 
(Grad-Shafranov) equation. Rutherford also showed that the island width $w$ grows with a linear time dependence in the early (small island) 
nonlinear phase. A quasilinear calculation predicted a further nonlinear slowing down of this growth \cite{Pellat}. 

The saturation of the tearing mode is a difficult issue, and its solution has been a stepwise process covering almost three decades. In 1977, a 
seminal work using a quasilinear calculation provided a first version of the formula $\Delta'(w_{sat})$  linking $\Delta'$ to the saturation 
amplitude $w_{sat}$ of the island width \cite{White}. In 1981, a new technique was introduced to deal with the case of a non vanishing current 
gradient $J'_{eq}(r_{s})$ on the resonant surface, and gave the correct expression for the leading order term in $w_{sat}$ of $\Delta'(w_{sat})$ 
in this case  \cite{Thya}. A later work provided a further term in this expression together with finite $\beta$ corrections \cite{Pletzer}. 
This term and finite $\beta$ corrections were also given in Ref. \onlinecite{Zakharov}. In 2004, the rigorous expression for $\Delta'(w_{sat})$ was 
provided for the case $J'_{eq}(r_{s}) = 0$ (Harris sheet pinch) by two different techniques, one of which uses explicitly the fact that the current 
profile is a flux function  \cite{MP}, while the other does not \cite{EO}. 

The aim of this paper is to provide a deeper insight in the nonlinear tearing mode by introducing, in particular, a powerful perturbation 
technique which does not use the fact that the current profile is a flux function. This makes possible, whenever useful, to avoid the assumption that 
plasma inertia, pressure or viscous effects are negligible, which opens a new route to deal with a background velocity profile and/or an external 
rotating forcing. Furthermore, whenever the current profile is a flux function, we show that the technique introduced in Ref. 
\onlinecite{Thya} can be simplified to make the derivation more direct, which divides by more than two the necessary algebra. As in all 
the above quoted works, $w_{sat}$ is assumed to be a small parameter in both of our techniques, and $\Delta'(w_{sat})$ is computed in a 
perturbative way for the case of a cylindrical geometry for any current gradient  $J'_{eq}(0)$.

Reference \onlinecite{Arcis} provides a short presentation of the new perturbative technique in slab geometry by the same authors. Our final 
formula for $\Delta'(w_{sat})$ depends on $w_{sat}$ through a term $w_{sat} \ln{(w_{sat}/ w_0)}$ where $w_0$ is a nonlinear scale 
length which was absent in previous work. We show that the nonlinear tearing mode comes with a modification of the background 
magnetic flux. It should also come with a modification of the resistivity profile. Indeed, since temperature is typically uniform on a 
given magnetic flux surface, so is the resistivity. Therefore, the development of a magnetic island is bound to flatten the resistivity profile. 
Since our RRMHD model does not incorporate the evolution of temperature, we provide the formulas for the saturation of the nonlinear 
tearing mode for both uniform and non uniform resistivity profiles. The same formula for $\Delta'(w_{sat})$ was obtained in parallel by 
another group who applied the technique of Ref. \onlinecite{Thya} to higher order \cite{HMP,IAEA} ; ref.\onlinecite{HMP} also 
provides expressions for $\Delta'(w_{sat})$ where the feedback of the magnetic island on the resistivity profile is accounted for in the case 
of a time independent thermal conductivity.

This paper is organized as follows. In Sec. \ref{model}, we introduce the main equations and we specify the different normalizations and 
notations adopted throughout the paper. The solutions of the (linearized) outer equations are given in Sec. \ref{outer}. We then tackle the 
problem of the saturated tearing mode in the nonlinear inner boundary layer (Sec. \ref{inner}), where the flux coordinate method, which 
exploits the fact that the current is a flux function is described first (Sec. \ref{constructive}). We outline how the calculations can be carried 
out consistently to the leading significant order in the island width expansion. This provides a benchmark for the more general perturbative 
approach, the main focus of this paper, which is presented in Sec. \ref{perturbative}. Section \ref{results} yields the modified Rutherford 
equation providing the whole nonlinear evolution of the island width, as well as the (lowest order) modification of the equilibrium magnetic 
flux function; it also provides a brief discussion of the validity limits of our method. Finally, Sec. \ref{conclusion} is devoted to the 
conclusions.

\section{DESCRIPTION OF THE MODEL}
\label{model}

\subsection{RMHD equations}

The RMHD equations are given by:
\begin{equation} \label{inertia1}
\partial_{t}\Delta_{\bot}\varphi+(\mathbf{v.\nabla})\Delta_{\bot}\varphi=\mathbf{B.\nabla}J+\nu \Delta_{\bot}^2
\varphi
\end{equation}
\begin{equation}\label{ohm1}
\partial_{t}\psi+\mathbf{B.\nabla}\varphi=\eta(J_{eq}-J)
\end{equation}
\begin{equation}\label{ampere1}
J=-\mu_{0}^{-1}
\Delta_{\bot}\psi
\end{equation}
where the mass density is uniform and taken equal to $1$ for simplicity, $\bot$ denotes the plane perpendicular to $\mathbf{e_{z}}$, 
$\psi$ is the poloidal flux (i.e. $\mathbf{B}\equiv B_{z}\mathbf{e_{z}}+\nabla\times (\psi\,\mathbf{e_{z}})$),  $\varphi$ is the 
stream function (i.e. $\mathbf{v}\equiv\mathbf{e_{z}}\times\nabla\varphi$), $\nu$ is the viscosity and $\eta$ is the resistivity. In 
the following, two models are considered: model A, in which the resistivity is uniform ($\eta=\eta_{A}$), and model B, in which it is 
not, but the electric field is ($\eta_{B}(r)J_{eq}(r)=E_{z}$). The equilibrium current profile $J_{eq}(r)$ then fully determines the plasma 
equilibrium which we have further assumed to be static ($\varphi_{eq}=0$).

\subsection{Single helicity perturbation of the equilibrium}

In the remainder of this paper, we work in cylindrical geometry and for now only consider the saturation of the single helicity 
$(m,n)$ perturbation of the equilibrium, to reintroduce time dependence at the very end of our calculations, which, as we shall see, can be 
done very simply. Therefore, all quantities henceforth depend on two variables only: $r$, and $\tau\equiv m\theta-nz/R $, where $R$ 
is the (simulated) major radius. Equations (\ref{inertia1}) and (\ref{ohm1}) then take the following form:
\begin{equation}
\label{inertia2}
B_{r}\partial_{r}J+\frac{nB_{z}}{R} \left( \frac{m}{nq}-1 \right)\partial_{\tau}J=\frac{m}{r}[\varphi,\Delta_{\bot}
\varphi]-\nu\Delta_{\bot}^2\varphi
\end{equation}
\begin{equation}\label{ohm2}
B_{r}\partial_{r}\varphi+\frac{nB_{z}}{R} \left( \frac{m}{nq}-1 \right)\partial_{\tau}\varphi=\eta(J_{eq}-J)
\end{equation}
where $q\equiv rB_{z}/RB_{\theta}$ is the safety factor, and, for any functions $f$ and $g$, $[f,g]\equiv \partial_{r}f\partial_
{\tau}g-\partial_{r}g\partial_{\tau}f$ is their 2D Jacobian or Poisson bracket.

\subsection{Helical flux function and final normalized equations}

It is easily seen that $\mathbf{B}.\nabla\psi\neq 0$, and, therefore, $\psi$ is not appropriate to describe magnetic surfaces. In contrast, 
the helical flux function $\psi^{*}=\psi+B_{z}nr^2/2mR$ verifies $\mathbf{B}.\nabla \psi^{*}= 0$, since $\mathbf{B}=B_{z}
\mathbf{h}+\nabla\times (\psi^{*}\mathbf{e_{z}})$, where $\mathbf{h}= \mathbf{e_{z}}+nr/mR\ \mathbf{e_{\theta}}$. We thus 
from now on work with $\psi^{*}$ instead of $\psi$. Finally, introducing the following normalizations
\begin{equation}\label{normalizations}
r=r_{0}\tilde{r}\mbox{ ; }\epsilon=\frac{r_{0}}{R}\mbox{ ; }J=J_{0}\tilde{J}\mbox{ ; }\psi^{*}=
\mu_{0}r_{0}^2J_{0}\widetilde{\psi}^{*}\mbox{ ; }\varphi=\frac{\eta_{0}}{\mu_{0}}\tilde{\varphi}\mbox{ ; }\eta=
\eta_{0}\tilde{\eta},
\end{equation}
where $r_{0}$ is the minor radius, $\mu_{0}r_{0}J_{0}= B_{z}$, $\eta_{0}=\eta_{A}$ for model A and $\eta_{0}=E_{z}/J_{0}$
for model B, equations (\ref{inertia2}) and (\ref{ohm2}) along with Ampere's law can be written in the following way:
\begin{equation}\label{inertia3}
\frac{m}{r}[J,\psi^{*}]=\frac{m}{rS^2}[\varphi,\Delta_{\bot}\varphi]-\frac{\Delta_{\bot}^2\varphi}{S.Re}
\end{equation}
\begin{equation}\label{ohm}
\frac{m}{r}[\varphi,\psi^{*}]=\eta(J_{eq}-J)
\end{equation}
\begin{equation}\label{ampere}
J=-\left(\frac{1}{r}\partial_{r}(r\partial_{r}\psi^{*})+\frac{m^2}{r^2}\partial_{\tau}^2\psi^*\right)+2\frac
{n\epsilon}{m}
\end{equation}
where we have omitted the " $\tilde{ }$ " for the sake of clarity. $S\equiv v_{A}r_{0}\mu_{0}/\eta_{0}$ is the Lundquist
number, 
$Re\equiv v_{A}r_{0}/\nu$ the Reynolds number, and $v_{A}\equiv J_{0}r_{0}\sqrt{\mu_{0}}$ the Alfv\'{e}n speed. Since $S\gg
1$
 and $Re \gg 1$, equation (\ref{inertia3}) merely gives \cite{Ruth}:
\begin{equation}\label{inertia}
[J,\psi^*]=0
\end{equation}
This means that $J=J(\psi^*)$, which is the basis of the flux coordinate method. Equations (\ref{ohm}), 
(\ref{ampere}) and (\ref{inertia}) are the basis of the following analytical work.

\subsection{Boundary conditions}

Equations (\ref{ohm}), (\ref{ampere}) and (\ref{inertia}) must be complemented with boundary conditions. Cylindrical geometry 
requires:
\begin{equation}\label{boundary0}
\lim_{r\rightarrow 0}\left(\frac{\partial_{\tau}\psi^*}{r}\right)=\left. \partial_{r}\psi^*\right|_{r=0}=0\ 
(\mathbf{B}_{\bot}(r=0)=\mathbf{0})\quad\mbox{and}\quad 
\lim_{r\rightarrow 0}\left(\frac{\partial_{\tau}\varphi}{r}\right)=\left. \partial_{r}\varphi\right|_{r=0}=0\ 
(\mathbf{v}_{\bot}(r=0)=\mathbf{0})
\end{equation}
Furthermore, since the plasma boundary is taken to be at $r=1$, the normal velocity component should vanish on this surface:
\begin{equation}\label{boundary1}
\left. \partial_{\tau}\varphi\right|_{r=1}=0
\end{equation}
As far as other boundary conditions for $\psi^*$ are concerned, they actually need not be specified explicitly to carry out the following 
calculations. Suitable boundary conditions are, for example, $\partial_{\tau}\psi^*|_{r=1}=0$, as in Ref. \onlinecite{White}, or
$\partial_{\tau}\psi^*|_{r=\infty }=0$, as in Ref. \onlinecite{Fitzpatrick}. 

This ends the definition of the differential problem to be solved.

\section{OUTER SOLUTION}
\label{outer}

\subsection{Outer equation}

As usual in tearing mode theory, we solve the differential problem as a boundary layer problem by 
matching an outer and an inner solution, and we approximate the outer solution by the linear ideal one. We set: $\psi^*=\psi^*_{eq}(r)+\delta^2
\psi_{1F}(r)\cos(\tau)+o(\delta^2)$. Linearizing (\ref{inertia})  yields:
\begin{equation}\label{outereq}
\psi_{1F}''+\frac{\psi_{1F}'}{r}+\left(\frac{J_{eq}'}{\psi_{eq}^{*'}}-\frac{m^2}{r^2}\right)\psi_{1F}=0
\end{equation}
From now on, we assume that $m\geq 2$, since we use the "constant-$\psi$" approximation which does not work for the $m=1$ mode. 
Equation (\ref{outereq}) then has two regular singular points: one at $r=0$ and the other at $r=r_{s}$, where $\psi^{*'}_{eq}(r_{s})=0$, 
or equivalently $q_{eq}(r_{s})=m/n$, since $q_{eq}=r\epsilon/(nr\epsilon/m-\psi_{eq}^{*'})$. $r_{s}$ is the location of the $(m,n)$ 
rational surface and, in the frame of the "constant-$\psi$" approximation, we can choose $\psi_{1F}(r_{s}) = 1$ so that $\delta$ is  the 
square root of the perturbation amplitude at $r=r_s$. The indicial equation for both points shows that, whenever the boundary conditions 
(\ref{boundary0}) are satisfied, $\psi_{1F}$ behaves like $r^m$ around $r=0$, and has a logarithmic singularity at $r=r_{s}$. This 
means that our perturbation expansion breaks down near the rational surface, which eventually has to be resolved thanks to a boundary 
layer centered upon it that we henceforth refer to as the "inner" region.

\subsection{Inner limit of the outer solution}

Let $\rho\equiv r-r_{s}$, we now give an expression for
$\psi_{1F}$ close to $\rho=0$. To do so, we first expand $J_{eq}$, $\psi_{eq}^*$ and 
$q_{eq}$ as $J_{eq}=\sum_{l\geq 0}a_{l}\rho^l$, $\psi_{eq}^*=\sum_{l\geq 2}b_{l}\rho^l$, and $q_{eq}=\sum_{l\geq 0}c_{l}
\rho^l$, where the $a's$, $b's$ and $c's$ are of course related to each other. In particular:
\begin{equation}\label{abc}
\begin{array}{|lc|l}
\displaystyle{b_{2}=\frac{n\epsilon}{m}-\frac{a_{0}}{2}} & & \displaystyle{c_{0}=\frac{m}{n}} \\
\displaystyle{b_{3}=-\frac{1}{3}\left(\frac{b_{2}}{r_{s}}+\frac{a_{1}}{2}\right)} & \mbox{and} & 
\displaystyle{c_{1}=\left(\frac{m}{n}\right)^2\frac{2b_{2}}{\epsilon r_{s}}} \\
\displaystyle{b_{4}=\frac{b_{2}}{4r_{s}^2}+\frac{a_{1}}{24r_{s}}-\frac{a_{2}}{12}} & & 
\displaystyle{c_{2}=\frac{1}{\epsilon r_{s}}\left(\frac{m}{n}\right)^2\left(3b_{3}-\frac{2b_{2}}{r_{s}}+
\frac{4mb_{2}^2}{n\epsilon r_{s}}\right)}
\end{array}
\end{equation}

Applying Froebenius' method to (\ref{outereq}) then allows us to derive the following expansion for $\psi_{1F}$:
\begin{eqnarray}\label{psi1}
\psi_{1F}&=&1+\frac{\Sigma'\pm\Delta'}{2}\rho+\left\{\alpha\frac{\Sigma'\pm\Delta'}{4}\left(1-\frac{1}
{\alpha r_{s}}\right)+\frac{1}{2}\left(\frac{m^2}{r_{s}^2}+\beta-\frac{\alpha^2}{2}(3-\frac{1}{\alpha r_{s}})\right)
\right\}\rho^2 \nonumber \\
& & +\alpha\left\{\rho+\frac{\alpha}{2}\left(1-\frac{1}{\alpha r_{s}}\right)\rho^2\right\}\ln|\rho|+O(\rho^3)
\end{eqnarray}
where $\pm\equiv sign(\rho)$, $\alpha=-a_{1}/2b_{2}$, and $\beta=-(a_{1}/4b_{2}r_{s}+a_{2}/b_{2}+a_{1}^2/8b_{2}^2)$. 
$\Delta'$ and $\Sigma'$ are two constants that are determined by the boundary conditions. They can be expressed as:
\begin{equation}\label{deltasigma}
\Delta'=\lim_{\epsilon\rightarrow 0^+}\left(\frac{\psi_{1F}'(r_{s}+\epsilon)-\psi_{1F}'(r_{s}-\epsilon)}{\psi_{1F}(r_{s})}
\right)\quad\mbox{and}\quad\Sigma'=\lim_{\epsilon\rightarrow 0^+}\left(\frac{\psi_{1F}'(r_{s}+\epsilon)+\psi_{1F}'
(r_{s}-\epsilon)}{\psi_{1F}(r_{s})}-2\alpha(1+\ln{\epsilon})\right\}
\end{equation}
$\Delta'$ is the usual tearing mode stability parameter \cite{Furth}. Note that, contrary to
$\Delta'$, $\Sigma'$ 
depends on normalization. Indeed, had we chosen to normalize $r$ with respect to, say, $\overline{r_{0}}$ instead of $r_{0}$, $\Sigma'$ 
would have been changed into $\overline{\Sigma'}=\Sigma'+2\alpha\ln|\overline{r_{0}}/r_{0}|$. This remark will prove to be 
important when we come to the saturation equation.

The logarithmic term appearing in (\ref{psi1}) implies that the perturbation expansion breaks down in a region centered on the rational 
surface. This comes from the fact that the quasilinear term in (\ref{inertia}) becomes of the same order as the linear one when $\rho$ is 
sufficiently small. Indeed, it is easy to see that:
\begin{equation} \label{breakdown}
[\psi_{eq},\delta^2J_{1F}\cos{\tau}]+[\delta^2\psi_{1F}\cos{\tau},J_{eq}]=O(\delta^2)\quad\mbox{and}\quad [\delta^2
\psi_{1F}\cos{\tau},\delta^2J_{1F}\cos{\tau}]=O(\frac{\delta^4}{\rho^2})
\end{equation}
where $ J_{1F}\cos{\tau}=-\Delta(\psi_{1F}\cos{\tau})$, and Eq. (\ref{breakdown}) immediately shows that the quasilinear term is 
no longer negligible when $\rho\sim\delta$. Therefore, in order to deal with the boundary layer, we use the stretched variable $\xi\equiv
\rho/\delta$. Since the outer solution $\psi_{eq}^*(r)+\delta^2\psi_{1F}(r)\cos(\tau)+o(\delta^2)$ is going to be matched to the 
inner solution which will be computed in the $\xi$ variable, we now re-write it in terms of $\xi$:
\begin{eqnarray}\label{psiout}
\psi_{out}^*(\xi,\tau)&=&\varsigma \delta^2\left\{|b_{2}|\xi^2+\varsigma  \cos{\tau}+\varsigma  \delta\left[b_{3}\xi^3
+\left(\frac{\Sigma'\pm\Delta'}{2}+\alpha(\ln|\xi|+\ln{\delta})\right)\xi\cos{\tau}\right] \right.  \nonumber \\
& & \left. +\varsigma  \delta^2\left[b_{4}\xi^4+\left(\alpha\frac{\Sigma'\pm\Delta'}{4}(1-\frac{1}{\alpha r_{s}})
+\frac{1}{2}(\frac{m^2}{r_{s}^2}+\beta-\frac{\alpha ^2}{2}(3-\frac{1}{\alpha r_{s}}))\right)\xi^2\cos{\tau}\right. 
\right. \nonumber \\
& & \left. \left. +\frac{\alpha ^2}{2}\left(1-\frac{1}{\alpha r_{s}}\right)\left(\ln|\xi|+\ln{\delta}\right)
\xi^2\cos{\tau}\right]\right\}
+o(\delta^4)
\end{eqnarray}
where $\varsigma  \equiv sign(b_2)$. Note the appearance of $\ln{\delta}$ terms that are due to the logarithmic singularity.

\section{INNER SOLUTION}
\label{inner}

\subsection{Inner equations}

Until otherwise stated, we now consider model A (constant resistivity) and will address model B later, where only minor modifications will
occur.
We work in $(\xi,\tau)$ variables. Matching with the outer solution (\ref{psiout}) implies that the inner one $\psi_{in}^*=O
(\delta^2)$, and we define a new function $\zeta$ as $\zeta\equiv \varsigma  \psi_{in}^*/\delta^2$. Assuming $\delta \xi \ll 1$, 
Eqs. (\ref{ohm}) and (\ref{ampere}) become:
\begin{equation}\label{ohmin}
\varsigma  \frac{m}{r_{s}}\delta\left(1-\delta\frac{\xi}{r_{s}}\right)[\zeta,\varphi]+o(\delta^2)=J-(a_{0}+a_{1}\delta
\xi+a_{2}\delta^2\xi^2)+o(\delta^2)
\end{equation}
\begin{equation}\label{amperein}
J=\frac{2n\epsilon}{m}-\varsigma \left\{\partial_{\xi}^2\zeta+\frac{\delta}{r_{s}}\left(1-\delta\frac{\xi}{r_{s}}
\right)\partial_{\xi}\zeta+\frac{m^2}{r_{s}^2}\delta^2\partial_{\tau}^2\zeta\right\}+o(\delta^2)
\end{equation}
where the Poisson bracket is now taken with respect to $(\xi,\tau)$ variables.

As regards the inertia equation, some care is needed before simply re-writing equation (\ref{inertia}) instead of (\ref{inertia3}). Indeed, 
we first have to compare the order of magnitude, {\it in the boundary layer}, of the different terms appearing in the latter equation before 
making any simplification. To do so, we need to have some information on $\varphi$. This can very simply be done by writing 
$\varphi=\delta^2\varphi_{1F}(r)\sin{\tau}+o(\delta^2)$ and solving (\ref{ohm}) to order $\delta^2$:
\begin{equation}\label{phiout}
\varphi_{1F}=\frac{\eta}{\psi_{eq}^{*'}}\left(\frac{m}{r}\psi_{1F}-\frac{\psi_{1F}'}{m}-\frac{r\psi_{1F}''}{m}\right)
\end{equation}
We see that, since $\varphi_{1F}$ goes as $\rho^{-2}$, then, in the boundary layer ($\rho\sim\delta$), $\varphi=O(1)$. It is now 
straightforward to show that the terms on the right hand side of (\ref{inertia3}) can be neglected if the following conditions hold:
\begin{equation}\label{conditioninertia}
\delta\gg S^{-2/5}\quad\mbox{and}\quad\delta\gg (S.Re)^{-1/6}
\end{equation}
The first of these is basically that $\delta$ be greater than the resistive layer width, a fact already pointed out in Ref. \onlinecite{Ruth}, and the 
second one is that it be larger than the visco-resistive length. Provided (\ref{conditioninertia})
is satisfied, the inertia equation can be written, in the inner domain, as in equation (\ref{inertia}):
\begin{equation}\label{inertiain}
[\zeta,J]=0
\end{equation}

In the following, we introduce two independent calculations of the nonlinear inner solution, which eventually has to be matched to 
(\ref{psiout}), using a perturbation expansion in $\delta$. The first one basically replaces equation (\ref{inertiain}) with $J(\xi,\tau)=
j(\zeta(\xi,\tau))$, i.e. it uses the fact that the current profile is a flux function. We only give a brief account of it, since it is an 
improvement of the technique described in Ref. \onlinecite{Thya}, and gives the same solutions as the second one which is new. The latter is indeed 
based on a classical perturbation expansion of all the involved functions, and is more flexible since it works also for problems where $J$ is 
not a function of $\psi$. 

\subsection{Flux coordinate method}
\label{constructive}

As already mentioned, equation (\ref{inertiain})  implies that $J(\xi,\tau)=j(\zeta(\xi,\tau))$. This makes natural 
the following change of variables:
\begin{equation} \label{change}
(\xi,\tau)\rightarrow (\zeta,\tau)
\end{equation}
In the $(\xi,\tau)$ plane, the curves $\zeta(\xi,\tau)=\, constant$ either cover the $-\pi\leq\tau\leq\pi$ interval or are closed. 
As a result, a given value of $\zeta$ may correspond to several values of $\xi$ for a given value of $\tau$ (at the maximum order of our 
calculations, there are only two values corresponding to the topology of a classical magnetic island). Therefore, the change of variable 
(\ref{change}) is one to one only in local domains. In such domains, we can solve $\zeta(\xi,\tau)=\, constant$ for $\xi$, which yields 
$\xi=X(\zeta,\tau)$. Consequently, $\varphi(\xi,\tau)$ becomes $\Phi(\zeta,\tau) = \varphi(X(\zeta,\tau),\tau)$. It follows that 
(\ref{ohmin}) can be recast into:
\begin{equation} \label{ohminchange}
\varsigma  \frac{m}{r_{s}}\left\{\partial_{\zeta}X\left(1+\delta\frac{X}{r_{s}}\right)\right\}^{-1}\partial_{\tau}
\Phi(\zeta,\tau)=j(\zeta)-J_{eq}(\delta X(\zeta,\tau))
\end{equation}
The periodicity in $\tau$ of $\Phi(\zeta,\tau)$ implies that the integration of 
(\ref{ohminchange}) on a flux surface covering the $-\pi\leq\tau\leq\pi$ interval, which we  refer to as $S_{\zeta}$, gives :
\begin{equation} \label{jzeta}
j(\zeta)=\int_{S_{\zeta}}J_{eq}(\delta X)\partial_{\zeta}X\left(1+\delta\frac{X}{r_{s}}\right)\, d\tau\left/
\int_{S_{\zeta}}\partial_{\zeta}X\left(1+\delta\frac{X}{r_{s}}\right)\, d\tau \right.
\end{equation}
A closed flux surface is described by a series of functions $X(\zeta,\tau)$ (two for the  maximum order of our calculations). These 
functions enable the generalization of Eq. (\ref{jzeta}) to closed flux surfaces by interpreting the integrals as loop integrals on a given flux 
surface. 

Equation (\ref{jzeta}) is the fundamental equation of the flux coordinate method. It was already derived in Ref. \onlinecite{Thya} for the case of a 
given resistivity profile (i.e. model B). The inner solution can be derived by combining it with a single differential equation, Eq. 
(\ref{amperein}), through the following iterative procedure. The first step sets $\delta = 0$ in Eq. (\ref{jzeta}). This yields $j_{0}=a_{0}$, 
which is set in Eq. (\ref{amperein}) to provide $\zeta_{0i}=|b_{2}|\xi^2 + a(\tau) \xi + b(\tau)$ where $a(\tau)$ and $b(\tau)$ are 
two unknown functions which are determined by matching with the outer solution (\ref{psiout}). Since the next order in the expansion
of $\zeta$ provided by  Eqs. (\ref{jzeta}) and (\ref{amperein}) is $\delta$, the matching brings to $a(\tau)$ and $b(\tau)$ terms of order 
$1$ and $\delta \ln{\delta}$. As a result, $\zeta_{0i}= \zeta_{0}+\varsigma \alpha\delta \ln{\delta} \, \xi \cos{\tau}$, 
where
\begin{equation} \label{z0}
\zeta_{0}=|b_{2}|\xi^2+\varsigma  \cos{\tau}.
\end{equation} 
Then the leading orders $X_{0i}$ of $X$ can
be computed by solving $\zeta =
\zeta_{0i}(X_{0i}(\zeta,\tau),\tau)$ at orders 1 and
 $\delta \ln{\delta}$ for $X_{0i}$. This yields
$X_{0i}(\zeta,\tau) =  X_0 -\varsigma \alpha \delta \ln{\delta} \cos{\tau}$, where $X_0 = \pm\sqrt{|b_2|^{-1}(\zeta -
\varsigma  \cos{\tau})}$. This calculation requires $|X_{0i}(\zeta,\tau)| \gg \delta \ln{1/ \delta}$, which excludes a small 
neighborhood of $\xi=0$.

This ends the first iteration of the calculation. We notice that two values of $\xi$ are related to one value of $\zeta$ at this level of 
approximation. The next iteration starts by setting $X_{0i}$ in Eq. (\ref{jzeta}), which brings orders $\delta$ and  $\delta^2 
\ln{\delta}$ to $j$. These orders are brought into Eq. (\ref{amperein}), which brings contributions of orders $\delta$ and $\delta^2 
\ln{\delta}$ to $\zeta$ which are completely defined by matching with the outer solution, and so on. The exclusion of a small 
neighborhood of $\xi=0$ is required at all orders. The calculation brings an expansion of $\Delta'$ in $\delta$ which is provided in the 
next section. We notice that each step of the flux coordinate method approximates $S_{\zeta}$ by its expression given by the available 
approximation of $X(\zeta,\tau)$.

In order to make the comparison with the perturbative method easier,  it is useful to notice that the results of the flux coordinate method 
provide perturbation expansions $\zeta=\sum_{l}\delta^{l}\zeta_{l}$, $j(\zeta)=\sum_{l}\delta^l j_{l}(\zeta) $, and 
$X(\zeta,\tau)=\sum_{l}\delta^l X_{l}(\zeta,\tau)$, where:
\begin{equation} \label{jzetal}
\begin{array}{|l}
\displaystyle{j_{0}=a_{0}} \\
\displaystyle{j_{1}=a_{1}\int_{S_{\zeta}}X_{0}\partial_{\zeta}X_{0}\, d\tau\left/
\int_{S_{\zeta}}\partial_{\zeta}X_{0}\, d\tau \right. }\\
\displaystyle{j_{2}=\int_{S_{\zeta}}\left\{\left(a_{2}+\frac{a_{1}}{r_{s}}\right)X_{0}^{2}\partial_{\zeta}X_{0}-
j_{1}\left(\partial_{\zeta}X_{1}+\partial_{\zeta}X_{0}\frac{X_{0}}{r_{s}}\right)+a_{1}\partial_{\zeta}(X_{0}X_{1})
\right\}\, d\tau\left/ \int_{S_{\zeta}}\partial_{\zeta}X_{0}\, d\tau \right. }.
\end{array}
\end{equation}
and
\begin{equation} \label{X01}
 X_{1}=-\,\partial_{\zeta}X_{0}\ \zeta_{1}(X_{0}(\zeta,\tau),\tau)
\end{equation}
The expression for $\zeta_{1}$ is given in the next section.

Notice that a differential equation for 
$X$ can be obtained by rewriting Ampere's law (\ref{amperein}) with respect to the new variables:
\begin{equation}\label{ampereflux}
J=\frac{2n\epsilon}{m}-\varsigma \left(\frac{1}{2}\left\{\partial_{\zeta}X^{-2}\right\}+\frac{\delta}
{r_s}\frac{1-\delta X/r_s}{\partial_{\zeta}X}+\frac{m^2}{r_s^2}\left\{\frac{1}{2}\partial_{\zeta}\left[\left(
\frac{\partial_{\tau}X}{\partial_{\zeta}X}\right)^2\right]-\partial_{\tau}\left[\frac{\partial_{\tau}X}
{\partial_{\zeta}X}\right]\right\}\right)+o(\delta^2)
\end{equation}

The flux coordinate method brings a series of simplifications to that of Ref. \onlinecite{Thya}: (i) no Ansatz is made about the solution, which 
brings only the non vanishing orders in $\delta$; (ii) the use of the $\zeta$ variable enables the same calculation to be formally done for 
flux surfaces inside and outside the magnetic island, and simplifies the calculation of flux surface averages; this divides the 
necessary algebra by more than a factor two; (iii) fewer quantities need to be defined to proceed with the calculation.

\section{Perturbative method}
\label{perturbative}

\subsection{Zeroth order}
The method is simply based on equations (\ref{ohmin}), (\ref{amperein}) and (\ref{inertiain}) along with the following natural 
expansions: $\zeta=\zeta_{0}+\delta\zeta_{1}+\delta^2\zeta_{2}+o(\delta^2)$, $J=J_{0}+\delta
J_{1}+\delta^2J_{2}+o(\delta^2)$ and $\varphi=\varphi_{0}+\delta\varphi_{1}+o(\delta)$, where a quantitiy with index $n$  
has an order smaller than $1/ \delta$ and larger or equal to 1. From (\ref{ohmin}), we directly obtain $J_{0}=a_{0}$, which, using 
(\ref{amperein}) and matching with (\ref{psiout}), immediately yields (\ref{z0}). We recognize the "constant-$\psi$" 
approximation that is valid to lowest order.

In the rest of the calculation, it will prove most useful to work in $(\zeta_{0},\tau\, ;\pm)$ variables where $\pm$ tells the sign of $\xi$. 
Let $f(\xi,\tau)$ be any function of the old variables, then we should introduce $\widehat{f}(\zeta_{0},\tau\, ;\pm)$ such that 
$\widehat{f}(\zeta_{0}(\xi,\tau),\tau\, ;\pm)=f(\xi,\tau)$. Nonetheless, to simplify formulas, we do not make that distinction in 
the following, which should always be kept in mind. In particular, $\xi$ should often be understood as $\xi=\pm\sqrt{|b_{2}|^{-1}
(\zeta_{0}-\varsigma \cos{\tau})}$.

Finally, we define $C_{x}$ contours as:
\begin{equation}\label{contour}
\left| \begin{array}{l} \mbox{if}\ x>1\quad
\mathcal{C}_{x}^{\pm}\equiv\{(\xi,\tau)\in \mathbb{R}^{\pm}\times
[-\pi,\pi] \,/\,\zeta_0(\xi,\tau)=x\} \\
\mbox{if}\ -1\leq x\leq 1\quad \mathcal{C}_x\equiv\{(\xi,\tau)\in
\mathbb{R}\times
[\frac{(\varsigma -1)\pi}{2}+\arccos{x},\frac{(\varsigma +3)\pi}{2}-\arccos{x}\,]\,/\,\zeta_0(\xi,\tau)=x\}
\end{array} \right.
\end{equation}
where the first line describes open curves and the second closed ones (from now on, we omit the $\pm$ superscript for open curves, 
which should not make any confusion). These contours merely represent lowest order magnetic surfaces.
In the following, we systematically make use of the fact that, for any single-valued and $\tau$-periodic function 
$f(\zeta_{0},\tau\, ;\pm)$:
\begin{equation}
\int_{C_{\zeta_{0}}}\!\!\!\partial_{\tau}f\, d\tau\equiv\left\langle \partial_{\tau}f \right\rangle=0
\end{equation}

\subsection{First order}

From (\ref{inertiain}), it can readily be seen that $J_{1}=j_{1}(\zeta_{0}\, ;\pm)$. We note, already, that we use the same notation 
$j_k$ as the one introduced in Sec. \ref{constructive}, since it will be shown shortly that they do refer to the same functions. Writing 
(\ref{ohmin}) to order $\delta$:
\begin{equation}\label{ohmin1}
-\varsigma \frac{m}{r_{s}}2|b_{2}|\xi\partial_{\tau}\varphi_{0}=a_{1}\xi-j_{1}
\end{equation}
and integrating the equation above along $C_{\zeta_{0}}$, we immediately derive the following expression for $j_{1}$:
\begin{equation}\label{j1}
j_{1}(\zeta_{0}\, ;\pm)=2\pi a_{1}H(\zeta_{0}-1)\left/ \left\langle \xi^{-1}\right\rangle \right.
\end{equation}
where $H$ is the Heaviside function. We now make an important remark concerning $j_1$. Indeed, expression (\ref{j1}) has a 
derivative singularity at $\zeta_0=1$, which is not physically acceptable. As already mentioned in Ref. \onlinecite{Thya}, 
this problem can be resolved thanks to a thin boundary layer centered around the separatrix $\zeta_0(\xi,\chi)=1$. However, contrary to 
what is claimed in Ref. \onlinecite{Thya}, it is not inertia but viscosity that is no longer negligible in equation (\ref{inertiain}) (see Appendix A). 
The current profile could thus, in principle, be regularized by solving the problem in this secondary visco-resistive boundary layer, using 
once again the technique of matched asymptotic expansions, a procedure similar to that already performed in Ref. \onlinecite{Edery} to 
regularize Rutherford's solution \cite{Ruth}. Nevertheless, that treatment needs not be done explicitly for present purposes. We 
therefore assume that $j_1$ is regular from now on, although we always use expression (\ref{j1}) in our calculations, which makes 
sense as long as it does not lead to divergences. Finally, using (\ref{j1}) together with (\ref{amperein}) provides $\zeta_{1}$:
\begin{equation}\label{z1}
\zeta_{1}=\frac{1}{2b_{2}}\left(\int_{1}^
{\zeta_{0}}j_{1}(x\, ;\pm)\, dx -\xi\int_{1}^{\zeta_{0}}\frac{j_{1}(x\, ;\pm)}{\xi(x,\tau\, ;\pm)}\, dx\right )-
\frac{|b_{2}|}{3r_{s}}\xi^3+A(\tau)\xi+B(\tau)
\end{equation}
where $A(\tau)$ and $B(\tau)$ are two unknown functions which have to be determined by the matching conditions.

Before proceeding to the matching with (\ref{psiout}), we note that, given any function $\widehat{f}(\zeta_{0}(\xi,\tau),\tau\, ;\pm)$, 
it is possible to obtain its asymptotic expansions in one of two ways: either express it as an explicit function of $(\xi,\tau)$ and directly 
derive its expansion as $|\xi|\gg 1$ (i.e. expand the related $f(\xi,\tau)$ function), or expand it as a function of $(\zeta_{0},\tau\, ;\pm)$ 
for $\zeta_{0}\gg 1$ and, only then, re-write that expansion with respect to $(\xi,\tau)$ while making $|\xi|\gg 1$ (see Appendix B). 
We use the latter method since it is much more convenient to implement.

It is straightforward to prove the following expansions:
\begin{equation}\label{expansions1}
\left\|\begin{array}{l}
\displaystyle{\xi(x,\tau\, ;\pm)=\frac{\pm 1}{\sqrt{2|b_{2}|}}\left(\sqrt{2x}-\varsigma \frac{\cos{\tau}}{\sqrt{2x}}+
O(x^{-3/2})\right)}  \\
\displaystyle{j_{1}(x\, ;\pm)=\frac{\pm a_{1}}{\sqrt{2|b_{2}|}}\sqrt{2x}+O(x^{-3/2})}  \\
\displaystyle{\int_{1}^{\zeta_{0}}j_{1}\, dx=\frac{\pm a_{1}}{3\sqrt{2|b_{2}|}}(2\zeta_{0})^{3/2}\pm\Omega+
O(\zeta_{0}^{-1/2})\   \mbox{where}\  
\Omega=\lim_{\zeta_{0}\rightarrow\infty}{\left(\pm\int_{1}^{\zeta_{0}}j_{1}\, dx-\frac{a_{1}}{3\sqrt{2|b_{2}|}}
(2\zeta_{0})^{3/2}\right)}} \\
\displaystyle{\int_{1}^{\zeta_{0}}\frac{j_{1}}{\xi}\, dx=a_{1}\zeta_{0}+\varsigma  a_{1}\ln{\sqrt{\zeta_{0}}}
\cos{\tau}-a_{1}+\Xi(\tau)+O(\zeta_{0}^{-1})\  \mbox{where}\   \Xi=\int_{1}^{\infty}\left(\frac{j_{1}}{\xi}-a_{1}-
\varsigma  a_{1}\frac{\cos{\tau}}{2x}\right)\, dx}
\end{array}\right.
\end{equation}
where $\Omega\equiv\omega a_{1}/\sqrt{2|b_{2}|}$, and $\omega$ is a numerical coefficient that is approximately equal to $-1.54$. 

Then, taking (\ref{expansions1}) and (\ref{z1}), we simply set $\zeta_{0}=|b_{2}|\xi^2+\varsigma \cos{\tau}$  and expand for 
$|\xi|\gg 1$:
\begin{equation}\label{z1expansion}
\zeta_{1}=-\frac{\varsigma }{3}\left(\frac{b_{2}}{r_{s}}+\frac{a_{1}}{2}\right)\xi^3+\varsigma \alpha\left(
\frac{\ln|b_{2}|}{2}+\ln|\xi|\right)\xi\cos{\tau}-\alpha\xi-\frac{\Xi(\tau)}{2b_{2}}\xi\pm\frac{\Omega}{2b_{2}}+
A(\tau)\xi+B(\tau)
+o(1)
\end{equation}
Since, in our approach, we have only taken into account the zeroth and first Fourier components from the outset, we should match only these 
two in our calculations. Thus, making use of (\ref{abc}), we can match (\ref{z1expansion}) with (\ref{psiout}), which determines 
$A(\tau)$ and $B(\tau)$
\begin{equation}\label{AB}
A(\tau)=\alpha-\varsigma \left(\frac{\alpha\ln|b_{2}|-\Sigma'}{2}-\varsigma \frac{\Xi_{1}}{2b_{2}}-\alpha\ln{\delta}
\right)\cos{\tau}\quad\mbox{and}\quad B(\tau)=0
\end{equation}
where $\Xi_{1}=\pi^{-1}\int_{-\pi}^{\pi} \Xi(\tau)\cos{\tau}\,d\tau$. What is important to note is that the first $\Delta'$ term 
appearing in (\ref{psiout}) cannot be matched with $\zeta_{1}$. Indeed, it would require the inclusion, in $A(\tau)$, of a quantity of the 
form $\pm\varsigma\Delta'\cos{\tau}/2$, which is not allowed since it would lead to $\zeta_{1}$'s being singular at the rational surface. 
Therefore, $\Delta'$ has to be matched with higher order terms, which will precisely provide the saturation condition we are looking for. 
Note also that, for the same reason, we cannot compensate the $\pm\Omega/2b_{2}$ term with $B(\tau)$. Since it is not matched in 
(\ref{psiout}) either, it implies a modification of the equilibrium magnetic flux that is of order $\delta^3$, which will be discussed later.

\subsection{Second order}
\label{perturbative-higher}

Moving on to the next order, it is easy to show that (\ref{inertiain}) now implies $J_{2}=j_{1}'(\zeta_{0}\, ;\pm)\zeta_{1}+j_{2}
(\zeta_{0}\, ;\pm)$. Here again, $j_{2}$ has to be determined through equation (\ref{ohmin}). To order $\delta^2$, we have:
\begin{equation}\label{ohmin21}
-\varsigma \frac{m}{r_{s}}2|b_{2}|\xi\left(\partial_{\zeta_{0}}\varphi_{0}\partial_{\tau}\zeta_{1}-
\partial_{\zeta_{0}}\zeta_{1}\partial_{\tau}\varphi_{0}-\partial_{\tau}\varphi_{1}+\frac{\xi}{r_{s}}\partial_{\tau}
\varphi_{0}\right)=j_{1}'\zeta_{1}+j_{2}-a_{2}\xi^2
\end{equation}
Besides, (\ref{ohmin1}) gives:
\begin{equation}\label{phi0}
-\varsigma \frac{m}{r_{s}}2|b_{2}|\partial_{\tau}\varphi_{0}=a_{1}-\frac{j_{1}}{\xi}\quad\mbox{and}\quad
-\varsigma  \frac{m}{r_{s}}2|b_{2}|\partial_{\zeta_{0}}\partial_{\tau}\varphi_{0}=\frac{j_{1}}{2|b_{2}|\xi^3}-
\frac{j_{1}'}{\xi}
\end{equation}
and, using (\ref{phi0}), (\ref{ohmin21}) can be recast into:
\begin{equation}\label{ohmin22}
-\varsigma  \frac{m}{r_{s}}2|b_{2}|\partial_{\tau}\left(\zeta_{1}\partial_{\zeta_{0}}\varphi_{0}-\varphi_{1}\right)+
j_{1}\left(\partial_{\zeta_{0}}\left[\frac{\zeta_{1}}{\xi}\right]-\frac{1}{r_{s}}\right)-a_{1}\partial_{\zeta_{0}}
\zeta_{1}+\left(\frac{a_{1}}{r_{s}}+a_{2}\right)\xi=\frac{j_{2}}{\xi}
\end{equation}
Then, integrating (\ref{ohmin22}) along $C_{\zeta_{0}}$ gives the expression for $j_{2}$
\begin{equation}\label{j2}
j_{2}=\left\langle\left(\frac{a_{1}}{r_{s}}+a_{2}\right)\xi+j_{1}\left(\partial_
{\zeta_{0}}\left[\frac{\zeta_{1}}{\xi}\right]-\frac{1}{r_{s}}\right)-a_{1}\partial_{\zeta_{0}}\zeta_{1}\right\rangle
\left/\left\langle \xi^{-1}\right\rangle \right.
\end{equation}
Note that $j_{2}$ turns out not to depend on $\pm$, i.e. is even in $\xi$. We also see that, to know $J_{2}$, we needed the full 
expression for $\zeta_{1}$. However, since we  stop the calculation at order 
$\delta^2$, only $\partial_{\xi}\zeta_{2}$ is now required to later perform the matching. We therefore integrate 
(\ref{amperein}) only once:
\begin{equation}\label{z2}
\partial_{\xi}\zeta_{2}=\frac{2|b_{2}|}{3r_{s}^2}\xi^3-\frac{j_{1}\zeta_{1}}{2b_{2}\xi}-\frac{\zeta_{1}}{r_{s}}
+\varsigma \frac{m^2}{r_{s}^2}\xi\cos{\tau}+\frac{1}{2b_{2}}\int_{\varsigma \cos{\tau}}^{\zeta_{0}}\!\!\!\left(
j_{1}\partial_{x}\left[\frac{\zeta_{1}}{\xi}\right]-\frac{j_{2}}{\xi} \right)\, dx +C(\tau)
\end{equation}
where, again, $C(\tau)$ has yet to be determined. 

Now that we know both the inner and outer solutions, we can proceed to the matching 
procedure, which is actually the most difficult part of the calculation. We begin by expanding the first terms in (\ref{z2}), which is easy to 
do since we have already derived (\ref{z1expansion}):
\begin{eqnarray}\label{z2expansion1}
\partial_{\xi}\zeta_{2}&=&\varsigma \left(4b_{4}+\frac{a_{2}}{3}+\frac{a_{1}}{6r_{s}}+\frac{a_{1}^2}{12b_{2}}
\right)\xi^3+\varsigma \left\{\alpha\left(1-\frac{1}{\alpha r_{s}}\right)\left(\frac{\Sigma'}{2}+\alpha(\ln|\xi|+
\ln{\delta})\right)-\frac{a_{1}b_{3}}{(2b_{2})^2}\right\}\xi\cos{\tau} \nonumber \\
& &+\varsigma \frac{m^2}{r_{s}^2}\xi\cos{\tau}\pm\alpha\frac{\Omega}{2b_{2}}\left(1-\frac{1}{\alpha r_{s}}\right)+
C(\tau)+\frac{1}{2b_{2}}\int_{\varsigma \cos{\tau}}^{\zeta_{0}}\!\!\!\left(j_{1}\partial_{x}\left[\frac{\zeta_{1}}
{\xi}\right]- \frac{j_{2}}{\xi} \right)\, dx
\end{eqnarray}
The last term of that expression is, therefore, the main part of the calculation. 

We first determine its diverging part. Making use of 
(\ref{j1}), (\ref{z1}) and (\ref{j2}), it is possible to derive the following expansions:
\begin{equation}\label{expansions2}
\left\|\begin{array}{l}
\displaystyle{j_{1}\partial_{\zeta_{0}}\left(\frac{\zeta_{1}}{\xi}\right)=\frac{\mp a_{1}}{3\sqrt{2|b_{2}|}}
\left(\frac{a_{1}}{2b_{2}}+\frac{1}{r_{s}}\right)\sqrt{2\zeta_{0}}\mp\frac{a_{1}^2\cos{\tau}}{\sqrt
{2\zeta_{0}}}(2|b_{2}|)^{-3/2}\mp\frac{ a_{1}\Omega}{4b_{2}\zeta_{0}}+O(\zeta_{0}^{-3/2})} \\
\displaystyle{j_{2}=\left(a_{2}+\frac{a_{1}}{6r_{s}}+\frac{a_{1}^2}{12b_{2}}\right)\frac{\zeta_{0}}{|b_{2}|}-\varsigma 
\frac{a_{1}\Omega}{\sqrt{2\zeta_{0}}}(2|b_{2}|)^{-3/2}+O(\zeta_{0}^{-1})} \\
\displaystyle{\frac{1}{2b_{2}}\int_{\varsigma \cos{\tau}}^{\zeta_{0}}\!\!\!\left(j_{1}\partial_{x}\left[
\frac{\zeta_{1}}{\xi}\right]-\frac{j_{2}}{\xi} \right)\, dx=} \\
\displaystyle{\quad \mp\varsigma \sqrt{\frac{\zeta_{0}}{|b_{2}|}}\left\{\left(\frac{a_{1}^2}{12b_{2}}+\frac{a_{2}}{3}+
\frac{a_{1}}{6r_{s}}\right)\frac{\zeta_{0}}{|b_{2}|}+ \left(\frac{7a_{1}^2}{24b_{2}^2}+\frac{a_{2}}{2b_{2}}+
\frac{a_{1}}{12b_{2}r_{s}}\right)\cos{\tau}\right\}+\mbox{converging term}}
\end{array}\right.
\end{equation}
Setting this result into (\ref{z2expansion1}) and expanding for $|\xi|\gg 1$ as usual, we finally get
\begin{eqnarray}\label{z2expansion2}
\partial_{\xi}\zeta_{2}&=&\varsigma  4b_{4}\xi^3-\varsigma \left\{\frac{3a_{1}^2}{8b_{2}^2}+\frac{a_{2}}{b_{2}}+
\frac{a_{1}}{4b_{2}r_{s}}-\frac{m^2}{r_{s}^2}-\alpha\left(1-\frac{1}{\alpha r_{s}}\right)\left(\frac{\Sigma'}{2}+
\alpha(\ln|\xi|+\ln{\delta})\right)\right\}\xi\cos{\tau} \nonumber \\
& & \pm\alpha\frac{\Omega}{2b_{2}}\left(1-\frac{1}{\alpha r_{s}}\right)+C(\tau)\pm\lim_{\zeta_{0}
\rightarrow\infty}\left\{\frac{1}{2b_{2}}\int_{\varsigma \cos{\tau}}^{\zeta_{0}}\!\!\!\left(j_{1}
\partial_{x}\left[\frac{\zeta_{1}}{|\xi|}\right]-\frac{j_{2}}{|\xi|} \right)\, dx \right. \nonumber \\
& &\left. +\varsigma \sqrt{\frac{\zeta_{0}}{|b_{2}|}}\left(\left[\frac{a_{1}^2}{12b_{2}}+\frac{a_{2}}{3}+\frac{a_{1}}
{6r_{s}}\right]\frac{\zeta_{0}}{|b_{2}|}+ \left[\frac{7a_{1}^2}{24b_{2}^2}+\frac{a_{2}}{2b_{2}}+\frac{a_{1}}{12b_{2}
r_{s}}\right]\cos{\tau}\right)\right\}+o(1)
\end{eqnarray}
Using (\ref{abc}) and the definition of $\beta$, it is possible to show that the first line of this equation matches all diverging 
terms of (\ref{psiout}). 

The last expression on the right hand side of (\ref{z2expansion2}) is the "converging term" mentioned in 
(\ref{expansions2}). In order for the matching to be complete, we have to select its first two Fourier components. We begin by the zeroth 
Fourier component
\begin{equation}\label{0Fourier1}
\pm\lim_{\zeta_{0}\rightarrow\infty}\left\{\frac{1}{4\pi b_{2}}\int_{-\pi}^{\pi}d\tau\int_
{\varsigma \cos{\tau}}^{\zeta_{0}}\!\!\!dx\,\left(j_{1}\partial_{x}\left[\frac{\zeta_{1}}{|\xi|}\right]-\frac{j_{2}}
{|\xi|}\right)+\varsigma \left(\frac{a_{1}^2}{12b_{2}}+\frac{a_{2}}{3}+\frac{a_{1}}{6r_{s}}\right)\left(
\frac{\zeta_{0}}{|b_{2}|}\right)^{3/2}\right\}
\end{equation}
We first note that 
\begin{equation}\label{0Fourier2}
\pm\int_{-\pi}^{\pi}d\tau\int_{\varsigma \cos{\tau}}^{\zeta_{0}}\!\!\!dx\,\left(j_{1}\partial_{x}\left[
\frac{\zeta_{1}}{|\xi|}\right]-\frac{j_{2}}{|\xi|} \right)=\int_{-\pi}^{\pi}d\tau\int_{\varsigma 
\cos{\tau}}^{\zeta_{0}}\!\!\!dx\,\left(\frac{j_{1}}{r_{s}}+a_{1}\partial_{x}\zeta_{1}-\left[a_{2}+\frac{a_{1}}
{r_{s}}\right]\xi\right)
\end{equation}
where use has been made of (\ref{j1}) and (\ref{j2}). We then derive the expansions below:
\begin{equation}\label{expansions3}
\left\|\begin{array}{l}
\displaystyle{\frac{-1}{4\pi b_{2}}\int_{-\pi}^{\pi}d\tau\int_{\varsigma \cos{\tau}}^{\zeta_{0}}\!\!\!dx\,
\xi=\mp\varsigma \frac{1}{3}\left(\frac{\zeta_{0}}{|b_{2}|}\right)^{3/2}+o(1)} \\
\displaystyle{\frac{1}{4\pi b_{2}r_{s}}\int_{-\pi}^{\pi}d\tau\int_{\varsigma \cos{\tau}}^{\zeta_{0}}\!\!\!dx\, j_{1}=
\pm \varsigma  \frac{a_{1}}{3r_{s}}\left(\frac{\zeta_{0}}{|b_{2}|}\right)^{3/2}\pm\frac{\Omega}{2b_{2}r_{s}}+o(1)} \\
\displaystyle{\frac{a_{1}}{4\pi b_{2}}\int_{-\pi}^{\pi}d\tau\int_{\varsigma \cos{\tau}}^{\zeta_{0}}\!\!\!dx\,
\partial_{x}\zeta_{1}=\mp\varsigma  \left(\frac{a_{1}}{6r_{s}}+\frac{a_{1}^2}{12b_{2}}\right)\left(\frac{\zeta_{0}}
{|b_{2}|}\right)^{3/2}\mp\alpha\frac{\Omega}{2b_{2}}+o(1)}
\end{array}\right.
\end{equation}
and thus see that the zeroth Fourier component exactly cancels out the $\pm\alpha(1-1/\alpha r_{s})\Omega/2b_{2}$ term on the 
second line of (\ref{z2expansion2}). The final matching condition is therefore given by $C(\tau)=0$ and
\begin{equation}\label{matching1}
\Delta'=\delta\lim_{\zeta_{0}\rightarrow\infty}\left\{\frac{\varsigma }{\pi b_{2}}\int_{-\pi}^{\pi}d\tau\cos{\tau}
\int_{\varsigma \cos{\tau}}^{\zeta_{0}}\!\!\!dx\,\left(j_{1}\partial_{x}\left[\frac{\zeta_{1}}{|\xi|}\right]-
\frac{j_{2}}{|\xi|}\right)+2\left(\frac{7a_{1}^2}{24b_{2}^2}+\frac{a_{2}}{2b_{2}}+\frac{a_{1}}{12b_{2}r_{s}}\right)
\sqrt{\frac{\zeta_{0}}{|b_{2}|}}\right\}
\end{equation}
The limit on the right hand side can be evaluated numerically, which 
eventually gives:
\begin{equation}\label{matching2}
\Delta'\simeq \delta\left\{1.64\left(\frac{a_{2}}{b_{2}\sqrt{2|b_{2}|}}+\frac{a_{1}\Sigma'}{4b_{2}\sqrt{2|b_{2}|}}+
\frac{a_{1}^2(\ln{\sqrt{|b_{2}|}}-\ln{\delta})}{(2|b_{2}|)^{5/2}}\right)+\frac{1.39\, a_{1}^2}{b_{2}^2\sqrt{2|b_{2}|}}+
\frac{0.65\, a_{1}}{r_{s}b_{2}\sqrt{2|b_{2}|}}\right\}+o(\delta)
\end{equation}
This equation is the final result of the matching procedure and will be re-written in a more pleasant form shortly. We see that 
(\ref{matching1}) requires the knowledge of divergences so as to substract them from the current profile integral. Furthermore, checking 
that the diverging terms do match correctly with the outer solution (\ref{psiout}) gives confidence in the validity of the saturation condition 
(\ref{matching2}). Lastly, given (\ref{jzetal}), (\ref{X01}) and the fact that $X_0(\zeta,\tau\, ;\pm)=\xi(\zeta,\tau\, ;\pm)$, it is 
easy to see that, setting $\zeta=\zeta_0+\delta \, \zeta_1+O(\delta^2)$, a simple Taylor expansion of (\ref{jzetal}) yields the same 
current profile functions $J_{n}$ that were derived above. In particular, the $j_k$'s introduced in Sec. \ref{constructive} and the ones derived 
in Sec. \ref{perturbative} are indeed the same functions, which justifies the use of the same notations, as was said previously. Therefore, 
the flux coordinate and perturbative methods do provide the same result, as already claimed.

\subsection{Non uniform resistivity}

We now treat model B (non uniform resistivity) using the perturbative method (it is easy to derive the equivalent of (\ref{jzeta}) for that 
model), and show that only slight differences occur while solving the inner equations. Of these, only 
Ohm's law is changed into
\begin{equation}
\label{ohminB}
\varsigma \frac{m}{r_{s}}\delta\left(1-\delta\frac{\xi}{r_{s}}\right)[\zeta,\phi]+o(\delta^2)= \eta(\delta\xi) J -1
\end{equation}
where $\eta(\rho)=\sum_{l\geq 0}d_{l}\rho^l$ satisfies $\eta J_{eq}=1$. Therefore, the $a$'s and $d$'s are easily related to each 
other and, in particular, $d_{0}=a_{0}^{-1}$, $d_{1}=-a_{1}/a_{0}^2$ and $d_{2}=a_{0}^{-2}(a_{1}^2/a_{0}-a_{2})$. 

The only difference then appears at order $\delta^2$, where $j_{2}$ now becomes:
\begin{equation}\label{j2B}
j_{2}^B=\left\langle\left(\frac{a_{1}}{r_{s}}+a_{2}-\frac{a_{1}^2}{a_{0}}\right)\xi+j_{1}
\left(\partial_{\zeta_{0}}\left[\frac{\zeta_{1}}{\xi}\right]-\frac{1}{r_{s}}+\frac{a_{1}}{a_{0}}\right)-a_{1}
\partial_{\zeta_{0}}\zeta_{1}\right\rangle\left/\left\langle \xi^{-1}\right\rangle \right.
\end{equation}
Using (\ref{expansions1}), it is straightforward to prove that $j_{2}^B$ has the same asymptotic expansion as that given in 
(\ref{expansions2}) for $j_{2}$. Consequently, the only differences that we have to determine are those coming from equations 
(\ref{0Fourier1}) and (\ref{matching1}). Making use of (\ref{expansions3}), one easily shows that:
\begin{equation}\label{0FourierB}
\lim_{\zeta_{0}\rightarrow\infty}\left\{\frac{1}{4\pi b_{2}}\int_{-\pi}^{\pi}d\tau\int_
{\varsigma \cos{\tau}}^{\zeta_{0}}\!\!\!\!\!\! dx\,\left(j_{1}\partial_{x}\left[\frac{\zeta_{1}}{|\xi|}\right]-
\frac{j_{2}^B}{|\xi|}\right)+\varsigma \left(\frac{a_{1}^2}{12b_{2}}+\frac{a_{2}}{3}+\frac{a_{1}}{6r_{s}}\right)
\left(\frac{\zeta_{0}}{|b_{2}|}\right)^{3/2}\right\}=\frac{\Omega}{2b_{2}}\left(\frac{1}{r_{s}}-\alpha-\frac{a_{1}}
{a_{0}}\right)
\end{equation}
Therefore, $\partial_{\xi}\zeta_{2}$ now has an asymptotic term of the form $\mp a_{1}\Omega/2b_{2}a_{0}$ which we cannot 
compensate with $C(\tau)$. It actually has to be matched with the order $\delta^3$ correction to the equilibrium magnetic flux that we 
mentioned earlier and which we will deal with somewhat later. It is interesting to point out that such a term was forbidden in 
model A because of current conservation during the relaxation (see Appendix C), and our calculations are thus consistent. 

As to the first Fourier harmonic, it can be shown to merely 
include a new term in (\ref{matching2}) whose coefficient can again be computed numerically  and is approximately equal to 
$-0.28\, \delta a_{1}^2/a_{0}b_{2}\sqrt{2|b_{2}|}$.

\section{RESULTS}
\label{results}

\subsection{Saturation equation}

The matching conditions that we have obtained for both models actually provide saturation equations for the island width, which is defined 
as $w\equiv 4\delta/\sqrt{2|b_{2}|}$ (i.e. it is the width of the separatrix of zeroth order magnetic surfaces as described by $\zeta_{0}$). 
We first need to reintroduce time dependence so as to make comparisons with previous results easier. Equation (\ref{ohmin}) is modified 
by adding $-\partial_{t}(\delta(t)^2\zeta)$ to the left hand side, whose lowest order term is $2\delta\dot{\delta}\cos{\tau}$, which 
changes (\ref{j1}) into 
\begin{equation}\label{j1bis}
j_{1}(\zeta_{0},t\, ;\pm)=\left(2\pi a_{1}H(\zeta_{0}-1)-2a_{0}\dot{\delta}\left\langle\frac
{\cos{\tau}}{\xi}\right\rangle\right)\left/ \left\langle \xi^{-1}\right\rangle \right.
\end{equation}
Since the new part added to $j_{1}$ is even in $\xi$, it contributes to the matching with 
$\Delta'$. We therefore take this new term into account and rewrite (\ref{matching2}) with respect to $w$ and for both models:
\begin{equation}\label{matching3}
a_{0}\dot{w}=1.22\, \Delta'+w\left\{\frac{1}{2}\alpha\left(\alpha\ln{w}+\frac{\Sigma'}{2}\right)-2.21\, \alpha^2+
0.40\, \frac{\alpha}{r_{s}}-\frac{a_{2}}{2b_{2}}-0.17\, \lambda\alpha\frac{a_{1}}{a_{0}}\right\}+o(w)
\end{equation}
where $\lambda=0$ for model A and $\lambda=1$ for model B. We now reintroduce normalizations explicitly and show that
\begin{equation}\label{abcurrent}
\left|\begin{array}{l}
\displaystyle{ \mathcal{A}\equiv \frac{\alpha}{ r_{0}}=\frac{ J_{eq}'(r_{s})}{J_{eq}(r_{s})}\left(1-\frac{2}{s}\right) }\\
\displaystyle{ 
\frac{\mathcal{B}}{2}\equiv -\frac{a_{2}}{2 r_{0}^2b_{2}}=\frac{ J_{eq}''(r_{s})}{2J_{eq}(r_{s})}\left(1-\frac{2}{s}\right) } \\
\displaystyle{a_{0}\frac{d}{d\tilde{t}}=\frac{\mu_{0}}{\eta_{eq}(r_{s})}r_{0}^2\frac{d}{dt}} \\
\end{array}\right.
\end{equation}
where $s\equiv r_{s}q_{eq}'(r_{s})/q_{eq}(r_{s})$ is the shear parameter, and use was made of the fact that $2b_{2}=a_{0}
s/(2-s)$, which can be proved using relations (\ref{abc}). Equation (\ref{matching3}), along with (\ref{abcurrent}), then provides 
the final evolution equation
\begin{equation}\label{matching4}
\frac{\mu_{0}}{\eta_{eq}(r_{s})}\frac{dw}{dt}=1.22\, \Delta'+w\left\{\frac{\mathcal{A}}{2}\left(\mathcal{A}
\ln{\frac{w}{r_{0}}}+\frac{\Sigma'}{2}\right)-2.21\, \mathcal{A}^2+0.40\, \frac{\mathcal{A}}{r_{s}}+
\frac{\mathcal{B}}{2}+\, 0.17\lambda\frac{\mathcal{A}^2s}{2-s}\right\}+o(w)
\end{equation}
where the $\Delta'$ and $\Sigma'$ are now to be understood as dimensional parameters whose definitions can trivially be deduced from 
(\ref{deltasigma}). A result similar to (\ref{matching4}) was obtained using a variant of Thyagaraja's technique \cite{HMP}, the only 
difference being the numerical coefficient of the $w\mathcal{A}/r_s$ term which was there found to be approximately equal to $0.22$.

We now make an important remark concerning (\ref{matching4}). Indeed, it can easily be shown that, despite the $\ln{w/r_{0}}$ term, 
it does not depend on the normalization length $r_{0}$. The reason for that is straightforward, and comes from the comment made below 
(\ref{deltasigma}) : $\Sigma'$ precisely depends on $r_{0}$ in such a way that the combination $\mathcal{A}\ln{w/r_{0}}+\Sigma'/2$ is 
actually normalization independent. It is therefore natural to define an intrinsic nonlinear scale length for the tearing mode as
\begin{equation}\label{scalinglength}
w_{0}\equiv r_{0}\exp{-\frac{\Sigma'}{2\mathcal{A}}}\quad\mbox{where}\quad
\Sigma'\equiv\lim_{\epsilon\rightarrow 0^+}\left(\frac{\psi_{1}'(r_s+\epsilon)+\psi_{1}'(r_s-\epsilon)}{\psi_{1}(r_{s})}-
2\mathcal{A}(1+\ln\frac{\epsilon}{r_{0}})\right\}
\end{equation}
and eventually rewrite (\ref{matching4}) in an explicit normalization independent way:
\begin{equation}\label{matchingfinal}
\frac{\mu_{0}}{\eta_{eq}(r_{s})}\frac{dw}{dt}=1.22\, \Delta'+w\left\{\frac{\mathcal{A}^2}{2}\ln{\frac
{w}{w_{0}}}-2.21\, \mathcal{A}^2+0.40\, \frac{\mathcal{A}}{r_{s}}+\frac{\mathcal{B}}{2}+0.17\, 
\lambda\frac{\mathcal{A}^2s}{2-s}\right\}+o(w)
\end{equation}

It is interesting to review past work in the light of this consideration.
First, one observes that the results obtained in Refs. \onlinecite{White} and \onlinecite{Zakharov} do depend on length normalization. 
The work done in Ref. \onlinecite{Thya} is formally normalization independent, since a change of normalization 
length induces a change in the saturation equation of order $\delta$, higher than the order  $\delta\ln{\delta}$ at which the calculations were stopped. 
However, the normalization issue strikes back in any practical application of the formula when an explicit value of $\delta$ must be set in 
the logarithm. Finally, the equation given in Ref. \onlinecite{Pletzer} turns out to be normalization independent, but lacks important order $\delta$ 
terms and does not provide any result for the symmetric case (i.e. $\mathcal{A}=0$) either. 

We finally mention that all the calculations which have been carried out so far can very easily be modified so as to fit the slab 
geometry case already treated in Ref. \onlinecite{Arcis}. One simply has to apply the following simplifications: 
\begin{equation}
a_{0}=1\, ,\quad b_{2}=-1/2\, , \quad \frac{m}{r_s}\rightarrow k\, , \quad r_{s}\rightarrow\infty\, , \quad s\rightarrow 
\infty\, , \quad \psi^*\rightarrow\psi\, , \quad\rho\rightarrow x\quad\mbox{and}\quad\tau\rightarrow \chi
\end{equation}
In particular, (\ref{matchingfinal}) immediately allows us to recover the evolution equation derived in Ref. \onlinecite{Arcis}.

\subsection{Modification of the equilibrium magnetic flux}

As pointed out when we derived the asymptotic behavior of the inner solution, there has to be an order $\delta^3$ modification of 
the equilibrium magnetic flux in the outer solution for the matching procedure to be complete, which is a result different from that obtained 
in Ref. \onlinecite{Pellat}, where it was found to be of order $\delta^4$. We therefore write $\psi^*=\psi^*_{eq}(r)
+\delta^2\psi_{1F}(r)\cos(\tau)+\delta^3\psi_{0F}(r)+o(\delta^3)$. From what we have done earlier, we know the following 
conditions that $\psi_{0F}$ must satisfy:
\begin{equation}\label{psi0conditions}
\lim_{r\rightarrow r_{s}}{\psi_{0F}(r)}=\pm\frac{\Omega}{2|b_{2}|}\quad\mbox{and}\quad
\lim_{r\rightarrow r_{s}}{\psi_{0F}'(r)}=\mp\lambda\frac{\Omega}{2|b_{2}|}\frac{a_{1}}{a_{0}}
\end{equation}
We now have to derive an equation for $\psi_{0F}$ and we will see that our simple perturbation technique is much easier to implement than the one
used in Ref. \onlinecite{Pellat} which is based on the $J=J(\psi)$ property. Setting $\varphi=\delta^2\varphi_{1F}(r)\sin{\tau}+\delta^3
\varphi_{0F}(r)+o(\delta^3)$, we see that equation (\ref{inertia}) is trivially satisfied to order $\delta^3$ and thus write (\ref{ohm}) 
at that same order, which merely gives $\psi_{0F}''+\psi_{0F}'/r=0$ and, consequently:
\begin{equation}\label{psi0}
\psi_{0F}(r)=D+E\ln{r}
\end{equation}

For model A (constant resistivity), it is clear that we simply have $\psi_{0F}^{A}=\pm\Omega/2|b_{2}|$, which does not violate the 
condition described in Appendix C. Physically, it does not lead to any change of the equilibrium magnetic field but merely reflects the fact 
that the saturation of the magnetic island leads to an increase of the (normalized) poloidal flux per unit length that, to lowest order, is equal 
to $2\delta^3\Omega/2|b_{2}|$. Making use of the normalizations (\ref{normalizations}) and integrating on the whole cylinder finally 
gives the total change of poloidal flux below:
\begin{equation}\label{fluxA} 
-0.048\, \mu_{0} R J_{eq}'(r_{s})w_{sat}^3
\end{equation}
where $w_{sat}$ is the saturated island width.

In the case of model B, things are not so straightforward. Indeed, in order for (\ref{psi0}) to match conditions (\ref{psi0conditions}), one 
would naively write
\begin{equation}\label{psi0Bnaive}
\psi_{0F}^{B}=\pm\frac{\Omega}{2|b_{2}|}\left(1-\frac{a_{1}r_{s}}{a_{0}}\ln{\frac{r}{r_{s}}}\right)
\end{equation}
The problem is that such a solution would not satisfy the boundary conditions of the problem given in (\ref{boundary0}) and would 
actually be singular at $r=0$. This means that the second condition in (\ref{psi0conditions}) is not acceptable. That problem can be solved 
by setting $C(\tau)=a_{1}\Omega/2b_{2}a_{0}$ in (\ref{z2}), which is perfectly allowed. $\psi_{0F}^{B}$ then finally becomes
\begin{equation}\label{psi0B}
\psi_{0F}^{B}=\frac{\Omega}{2|b_{2}|}\left(\pm 1 -2H(r-r_{s})\frac{a_{1}r_{s}}{a_{0}}\ln{\frac{r}{r_{s}}} \right)
\end{equation}
which results in the poloidal flux being changed into
\begin{equation}\label{fluxB} 
-0.048\, \mu_{0} R J_{eq}'(r_{s})w_{sat}^3\left(1-\frac{r_{s}J_{eq}'(r_{s})}{J_{eq}(r_{s})}\ln{\frac{r_{0}}
{r_{s}}}\right).
\end{equation}
This change of order $\delta^3$ in the flux is a natural consequence of $J_0 = a_0$ and of the expression (\ref{j1}) for $j_1$. Indeed, 
they tell that, in the nonlinear regime, the current has a plateau inside the island. Since the width of the island is $O(\delta)$, this brings a 
change $O(J_{eq}'(r_{s}) \delta^2)$ to the magnetic field and $O(J_{eq}'(r_{s}) \delta^3)$ to the magnetic flux.

\subsection{Validity limits of the method}

Since we have used a perturbation expansion in $\delta$, the first condition that should be met for our calculations to be valid is obviously 
$\delta\ll 1$. For instance, in the case when $\mathcal{A}=0$ and $\mathcal{B}\geq 0$, equation (\ref{matchingfinal}) would 
predict exponential growth of the island width. However, since our result is no longer valid when $\delta$ approaches unity, this does not 
necessarily mean that the system would lead to a disruption, for there might be saturation with a large island.

The second, more limiting condition is that $\Delta'$ be not too large, which we shall make more precise right away. Suppose that 
$\Delta'$ is such that $\Delta'\delta/2\sim 1$. Then, in (\ref{psiout}), the lowest order term should include the $\Delta'\delta$ one 
and, therefore, $\zeta_{0}$ would become
\begin{equation}
\zeta_{0}=|b_{2}|\xi^2+\varsigma \left(1\pm\frac{\Delta'\delta}{2}\xi\right)\cos{\tau}
\end{equation}
However, this would not be allowed, because of the singularity at $r=r_{s}$, and our method would thus lead to a dead end.

This basic analysis is in good agreement with recent numerical results obtained by Loureiro {\it et al.} \cite{Loureiro} for the symmetric 
tearing mode in slab geometry. Indeed, since, in that case, the island width is simply given by $w=4\delta$, the condition 
$\Delta'\delta/2\sim 1$ gives $w\Delta'\sim 8$, which is surprisingly close to the condition derived in Ref. \onlinecite{Loureiro} 
($w\Delta'\simeq 8.2$). When this condition is met, Loureiro {\it et al.} observe the formation of current sheets, which, typically, means 
that the first Fourier harmonic of the perturbation is no longer dominant with respect to higher ones, which contradicts the "constant-$\psi$" 
approximation used in our approach.

\section{CONCLUSION}
\label{conclusion}

We have provided a rigorous solution to the simple tearing mode problem in cylindrical geometry using both the flux coordinate method and 
a new perturbation technique, and our calculations can directly be transposed to the case of a plasma slab. The final evolution equation 
contains all terms of order $w$ and has been explicitly shown to be normalization independent, a necessary physical requisite. We have also 
shown that the saturation of the tearing mode leads to a modification of the equilibrium magnetic flux function which we have been able to 
fully determine and which is  consistent with the condition of current flux conservation in the case of a uniform resistivity profile. Lastly, we 
have discussed the limits of validity of our approach and we have derived a qualitative condition on $\Delta'$ which is in good agreement with the 
recent numerical study carried out in Ref. \onlinecite{Loureiro}.

Besides the actual results shown here, it is important to appreciate the importance of establishing solid analytic techniques for future work. In 
particular, the perturbation method, which does not rely on a functional dependence between the current and the flux function, is rather 
promising for the treatment of more general models than conventional reduced MHD. For instance, the forced tearing mode in rotating 
plasmas is currently being revisited and a first application to the static case was given in Ref. \onlinecite{Arcis}. In the longer term, two-fluid 
models with diamagnetic effects should also be accessible to analytic investigations with these techniques.

\begin{acknowledgments}

We acknowledge fruitful discussions with J. Hastie, F. Militello and F. Porcelli that led to a joint oral contribution at the last IAEA meeting 
\cite{IAEA}, and thank F. Militello for the comparison between the numerical coefficients of the island width evolution equation. We also 
thank A.I. Smolyakov for pointing out Ref. \onlinecite{Pellat}. One of us (DFE) thanks Y. Elskens for useful comments about boundary 
conditions in asymptotic matching techniques.

\end{acknowledgments}

\appendix

\section{VISCOUS BOUNDARY LAYER AROUND THE SEPARATRIX}

We want to evaluate the first neglected terms in (\ref{inertiain}) and derive their behavior around the separatrix $\zeta_0=1$. Equation 
(\ref{ohmin1}) gives:
\begin{equation}\label{phi0in}
\varphi_0=\frac{r_s}{2mb_2}\left(\int_0^{\tau}\frac{j_1}{\xi}\, dy-a_1\tau\right)+\Phi_0(\zeta_0)
\end{equation}
Furthermore, using (\ref{phiout}), the matching condition for $\varphi_0$ is
\begin{equation}\label{phimatching}
\varphi_0=-\frac{\alpha r_s}{2mb_2}\times \frac{\sin{\tau}}{\xi^2}+o(\xi^{-2})
\end{equation}
which is automatically satisfied by the first term on the right hand side of (\ref{phi0in}).

To lowest order, the corrections to (\ref{inertiain}) are given by:
\begin{equation}\label{inertiaphi}
-\varsigma \delta^2 [\zeta_0,j_1]\sim \frac{1}{\delta^3S^2}[\varphi_0,\partial^2_{\xi}\varphi_0]-\frac{r_s}{mS.Re
\delta^4}\partial^4_{\xi}\varphi_0
\end{equation}
We thus see that, if the magnetic Prandtl number $S/Re$ is of order unity, the main correction is due to viscosity and we can neglect inertia. 
Integrating (\ref{inertiaphi}) on $C_{\zeta_0}$ then gives the following condition:
\begin{equation}\label{condition}
\partial^2_{\zeta_0}\left\langle\xi^3\partial^2_{\zeta_0}\varphi_0\right\rangle=\partial^2_{\zeta_0}\left(\left
\langle \xi^3\right\rangle\partial^2_{\zeta_0}\Phi_0\right)+\frac{r_s}{2mb_2}\partial^2_{\zeta_0}\left(\left\langle
\xi^3\int_0^{\tau}\partial^2_{\zeta_0}\left\{\frac{j_1}{\xi}\right\}\, dy\right\rangle\right)=0
\end{equation}
Since it is easy to prove that $\displaystyle{\left\langle\xi^3\int_0^{\tau}\partial^2_{\zeta_0}\left\{\frac{j_1}{\xi}
\right\}\, dy\right\rangle=0}$, we simply have $\Phi_0^{''}=0$. Therefore, the main neglected term in (\ref{inertiain}) is given by:
\begin{equation}\label{correction}
\frac{4b_2^2}{S.Re \delta^6}\left(\frac{r_s}{m}\right)^2\partial^2_{\zeta_0}\left(\xi^3\int_0^{\tau}\partial^2_
{\zeta_0}\left\{\frac{j_1}{\xi}\right\}\, dy\right)
\end{equation}
It can be shown that, close to the separatrix, $j_1$ behaves as
\begin{equation}
j_1\sim\frac{2\pi a_1}{\sqrt{2|b_2|}}(\ln{32}-\ln{(\zeta_0-1)})^{-1}+O(\zeta_0-1)
\end{equation}
If the position is not too close to the O-point or to the X-point, it is then easy to see that 
\begin{equation}
\partial^2_{\zeta_0}\left(\xi^3\int_0^{\tau}\partial^2_{\zeta_0}\left\{\frac{j_1}{\xi}\right\}\, dy\right)=
O\left(\frac{1}{(\zeta_0-1)^4}\right)
\end{equation}
This means that (\ref{correction}) is no longer negligible when the following condition holds:
\begin{equation}\label{viscousboundary}
\zeta_0-1\sim \left(\frac{\delta_v}{\delta}\right)^{3/2}
\end{equation}
where $\delta_v\equiv (S.Re)^{-1/6}$ is the visco-resistive length. (\ref{viscousboundary}) then determines the size of the viscous 
boundary layer centered on the separatrix which should, in principle, regularize all the singularities that appear in the current profile.

\section{VALIDITY OF THE $\zeta_{0}\rightarrow\infty$ ASYMPTOTIC MATCHING}

We first define the equivalence relation below:
\begin{equation}\label{equivalence}
\mbox{Let}\ (f,g)\in \left(\mathbb{R}^{\mathbb{R}}\right)^2,\ f\stackrel{\pm\infty}{\sim_{x}}g
\Leftrightarrow \lim_{x\rightarrow\pm\infty}\left(f(x)-g(x)\right)=0
\end{equation}
If we define the function $\widehat{f}(\zeta_{0},\tau\, ;\pm)$ such that $\widehat{f}(\zeta_{0}(\xi,\tau),\tau\, ;\pm)=
f(\xi,\tau)$, what we want to show is the following property:
\begin{equation}\label{theorem}
\mbox{If}\ f\stackrel{\pm\infty}{\sim_{\xi}}f_{\infty}\ \mbox{and}\ \widehat{f}\stackrel{+\infty}{\sim_
{\zeta_{0}}}\widehat{f}_{\infty}\ \mbox{then}\ \widehat{f}_{\infty}(\zeta_{0}(\xi,\tau),\tau\, ;\pm)
\stackrel{\pm\infty}{\sim_{\xi}}f_{\infty}(\xi,\tau)
\end{equation}
Let $\varepsilon\in\mathbb{R}^{+*}$, then, by definition of the equivalence relation (\ref{equivalence}), we have:
\begin{equation}\label{fbar}
\exists\, \widehat{M}_{\varepsilon}\in\mathbb{R}^{+},\ \forall (\zeta_{0},\tau)\in\mathbb{R}^2,\ 
\zeta_{0}\geq\widehat{M}_{\varepsilon}\Rightarrow |\widehat{f}(\zeta_{0},\tau\, ;\pm)-\widehat{f}_{\infty}
(\zeta_{0},\tau\, ;\pm)|\leq\varepsilon
\end{equation}
Now, let $M_{\varepsilon}=\sqrt{|b_{2}|^{-1}(\widehat{M}_{\varepsilon}+1)}$, then it is clear that:
\begin{equation}\label{f}
\forall (\xi,\tau)\in\mathbb{R}^2,\ |\xi|\geq M_{\varepsilon}\Rightarrow |\widehat{f}(\zeta_{0}(\xi,\tau),\tau\, ;\pm)
-\widehat{f}_{\infty}(\zeta_{0}(\xi,\tau),\tau\, ;\pm)|\leq\varepsilon
\end{equation}
Since $\widehat{f}(\zeta_{0}(\xi,\tau),\tau\, ;\pm)=f(\xi,\tau)$, we immediately deduce (\ref{theorem}).

\section{CURRENT FLUX CONSERVATION IN THE CASE OF UNIFORM RESISTIVITY}

In model A, Ohm's law is written as:
\begin{equation}\label{ohmA}
\frac{m}{r}[\varphi,\psi^{*}]=J_{eq}-J
\end{equation}
Integrating the left hand side on a poloidal section $S=\{(r,\tau)\in [0,1]\times [0,2\pi]\}$ gives:
\begin{equation}
\int\!\!\!\int_{S} \frac{m}{r}[\varphi,\psi^{*}]\, dx\, dy=-\int_{0}^1dr\int_{0}^{2\pi}d\tau\left\{
\partial_{r}(\psi^*\partial_{\tau}\phi)-\partial_{\tau}(\psi^*\partial_{r}\phi)\right\} \nonumber
\end{equation}
\begin{equation}\label{lhs}
\quad =\int_{0}^{2\pi}d\tau\left\{\psi^*(0,\tau)\partial_{\tau}\phi|_{r=0}-\psi^*(1,\tau)
\partial_{\tau}\phi|_{r=1}\right\}+\int_{0}^1dr\left\{\psi^*(r,2\pi)\partial_{r}\phi|_{\tau=2\pi}-\psi^*(r,0)
\partial_{r}\phi|_{\tau=0}\right\}
\end{equation}
where we have used the Green-Riemann theorem. It is clear that, because of periodicity and of the boundary conditions (\ref{boundary1}), 
the last line of (\ref{lhs}) is equal to zero and, therefore, integrating the right hand side of (\ref{ohmA}) gives:
\begin{equation}\label{conditionA}
\int\!\!\!\int_{S}(J_{eq}-J)\, dx\, dy=\int_{\partial S}(\mathbf{B_{eq}}-\mathbf{B}).d\mathbf{l}=
\frac{1}{m}\left(\int_{0}^{2\pi}d\tau\partial_{r}\psi^*|_{r=1}-2\pi \psi_{eq}^{*'}(r=1)\right)=0
\end{equation}
where we have used Stokes' theorem and again the periodicity in $\tau$. Condition (\ref{conditionA}) thus imposes that $\partial
_{r}\psi_{0F}|_{r=1}=0$.


\begin{thebibliography}{30}

\bibitem{Furth}
H.P. Furth, J. Killeen, M.N. Rosenbluth, Phys. Fluids {\bf 6}, 459 (1963).

\bibitem{Coppi} 
B. Coppi, Physics Letters {\bf 11}, 226 (1964).

\bibitem{LP} 
G. Laval and R. Pellat, C. R. Acad. Sci. Paris {\bf 259}, 1706 (1964).

\bibitem{LPV}
 G. Laval, R. Pellat and M. Vuillemin, {\it Plasma Physics and Contr. Nucl. Fusion Research}, IAEA, Vienna, Vol. II, 736 (1966)

\bibitem{Tebaldi}
C. Tebaldi and M. Ottaviani, J. Plasma Phys. {\bf 62}, 513 (1999).

\bibitem{Dahlburg}
This fact is also implicit in: R.B. Dahlburg, T.A. Zang, D. Montgomery and M.Y. Hussaini, {\it Proceedings of the National Academy of 
Sciences of the United States of America} (1983), Vol. 80, No. 18, p. 5798.

\bibitem{Luce}
T.C. Luce, M.R. Wade, J.R. Ferron, P.A. Politzer, A.W. Hyatt, A.C.C. Sips and M. Murakami, Phys. Plasmas {\bf 11}, 2627 (2004).

\bibitem{Strauss}
H.R. Strauss, Phys. Fluids {\bf 19}, 134 (1976).

\bibitem{Ruth}
P. Rutherford, Phys. Fluids {\bf 16},  1903  (1973).

\bibitem{Pellat}
R. Pellat, M. Frey and M. Tagger, J. Physique {\bf 45}, 1615 (1984).

\bibitem{White}
R.B. White, D.A. Monticello and Marshall N. Rosenbluth,  Phys. Fluids {\bf 20}, 800 (1977).

\bibitem{Thya}
A. Thyagaraja, Phys. Fluids {\bf 24}, 1716 (1981).

\bibitem{Pletzer}
A. Pletzer and F.W. Perkins, Phys. Plasmas {\bf 6}, 1589 (1999).

\bibitem{Zakharov}
L.E. Zakharov, A.I. Smolyakov and A.A. Subbotin, Sov. J. Plasma Physics {\bf 16}, 451 (1990).

\bibitem{MP}
F. Militello  and F. Porcelli, Phys. Plasmas  {\bf 11}, L13 (2004).

\bibitem{EO}
D.F. Escande and M. Ottaviani, Phys. Lett. A {\bf 323}, 278 (2004).

\bibitem{Arcis}
N. Arcis, D.F. Escande and M. Ottaviani, Phys. Lett. A {\bf 347}, 241 (2005)

\bibitem{HMP}
R. J. Hastie, F. Militello  and F. Porcelli, Phys. Rev. Lett. {\bf 95}, 065001 (2005)

\bibitem{IAEA}
R. J. Hastie, F. Militello,  F. Porcelli, N. Arcis, D.F. Escande, and M. Ottaviani, {\it Proceedings of the 20th IAEA Fusion Energy 
Conference}, Vilamoura, Portugal, edited by Marianne Spak (IAEA, Vienna, 2004), PD/1-1

\bibitem{Fitzpatrick}
R. Fitzpatrick, Nucl. Fusion {\bf 33}, 1049 (1993).

\bibitem{Edery}
D. Edery, M. Frey, J.P. Somon, M. Tagger, J.L. Soule, R. Pellat and M.N. Bussac, Phys. Fluids {\bf 26} (5), 1165 (1983).

\bibitem{Loureiro}
N.F. Loureiro, S.C. Cowley, W.D. Dorland, M.G. Haines and A.A. Schekochihin, Phys. Rev. Lett. {\bf 95}, 235003 (2005).

\bibitem{Waelbroeck}
F.L. Waelbroeck, Phys. Fluids B {\bf 1} (12), 2372 (1989).


\end{thebibliography}
\end{document}